\documentstyle[aps,prb,floats,epsf,amsbsy]{revtex}
\begin{document}
\draft
\def\overlay#1#2{\setbox0=\hbox{#1}\setbox1=\hbox to \wd0{\hss #2\hss}#1
-2\wd0\copy1}
\newcommand{\ve}[1]{\boldsymbol{#1}}

\twocolumn[\hsize\textwidth\columnwidth\hsize\csname@twocolumnfalse\endcsname
\title{Peierls Instabilities in Quasi-One-Dimensional 
Quantum Double-Well Chains}
\author{Natalie I. Pavlenko\\
Institute for Condensed Matter Physics\\
1, I. Svientsitsky Str., UA-290011, Lviv, Ukraine}
\date{\today}
\maketitle
\begin{abstract}
Peierls-type instabilities in quarter-filled ($\bar{n}=1/2$) and
half-filled ($\bar{n}=1$) quantum double-well hydrogen-bonded chain
are investigated analytically in the framework of two-stage 
orientational-tunnelling model with additional inclusion of the interactions 
of protons with two different optical phonon branches. It is shown that 
when the energy of proton-phonon coupling becomes large, the system 
undergoes a transition
to a various types of insulator states. The influence of two different 
transport amplitudes on ground states properties is studied. The results 
are compared with the pressure effect experimental investigations in 
superprotonic systems and hydrogen halides at low temperatures.
\end{abstract}


\vskip2pc]
\narrowtext

\section{INTRODUCTION}

It is long been known that the formation of hydrogen bonds between 
molecules or ionic groups is responsible for a drastic changes in
a wide variety of entire system properties such as structural phase
transformations and proton ordering phenomena \cite{blinc,aksenov}. In 
addition, proton transport phenomena in H-bonded materials and
superionic properties discovered in some hydrogen-bonded
crystals (for example, M$_3$H(AO$_4$)$_2$ class where 
M=Rb, Cs, NH$_4$; A=Se, S) are related closely to hydrogen-bonded network
rearrangement. On heating these crystals transform into superionic 
conducting phase with statistically disordered hydrogen-bonded network 
(Fig.~1(a)).
The protons can migrate through the two-dimensional conducting planes
with low activation energy ($\sim 0.1$~eV). In this case protonic 
conductivity increases significantly to the value about 0.1~$\Omega^{-1} 
\cdot$~cm$^{-1}$. It is generally accepted \cite{belushkin} that the 
two-stage conduction mechanism is required to sustain proton transport. The 
intrabond proton tunnelling along the hydrogen bridge is connected with the 
transfer of ionic positive and negative charged defects, whereas the 
intermolecular proton transfer
due to reorientations of molecular group with proton leads to the breaking
of the hydrogen bond and creation of a new one between another pair of
molecular complexes. It should be noted that the formation of the hydrogen
bridge induces the distortion of groups involved in 
hydrogen bond towards the proton that results in the shortening of the 
bond \cite{pietraszko}. By this means the 
protonic polaron is localized between distorted ionic groups in the 
low-temperature ferroelastic phases, giving rise in this case to the 
dimerized structure. As has been shown in Ref.~\onlinecite{pavlenko}, 
the small-radii 
polaron is formed due to the strong coupling of proton with optical 
stretching vibration modes of the oxygen ions. It is evident that such an 
transformations from the superionic phase occurring in systems on cooling 
have the mixed (displacive and order-disorder) character. 
Theoretical investigations of various
ferroelectric-type orderings in hydrogen-bonded systems have been
based generally on pseudospin Ising-type models with additional including 
of the pseudospin-phonon interactions to describe the coupling of protons 
with lattice vibration modes. In particular, the quantum 
double-well chain with quartic symmetric double-well potential has been 
used to model the transition from the symmetry-broken to the 
symmetry-restored ground state in hydrogen halides 
HX (X=F, Br, Cl) \cite{wang}
which consist of hydrogen-bonded chains with weak 
interchain coupling. The dynamics of both ionic and orientational defects
created by the rotations of molecular groups in hydrogen halides has been
studied in the framework of classic approach based on soliton model 
\cite{savin}.

It must be emphasized that taking into account the two-stage transport 
mechanism renders the pseudospin formalism unsuitable for the proton 
subsystem description since the number of protons can differ from the 
number of possible (virtual) hydrogen bonds and the proton occupancy of 
each bond in principle can be other than unity due to reorientational 
hopping and consequent feasibility of proton migration along the chain. 
Such a situation, as an example, is observed in superionic materials of 
M$_3$H(AO$_4$)$_2$-type which transform on cooling into dielectric state 
with dimerized structure \cite{pietraszko2}. 
It should be noted that such type
of transitions to dielectric states is 
reminiscent of that of electronic systems in which the Peierls 
instabilities are observed. There have been many works to study the 
metal-insulator Peierls transitions in electron-phonon systems which are 
unstable against the electron-phonon interactions \cite{peierls,rise}. It 
is common knowledge that the Peierls instabilities occur with the 
formation of Peierls gap at $\ve{k}=\pm \ve{k}_F$ ($\ve{k}_F$ is the Fermi 
level)in the electronic energy band that connected with the electronic 
charge density waves condensation and structural lattice distortion 
modulations with $\ve{q}=2\ve{k}_F$. The 
appearance of insulator state together with the structural transformation 
can be modelled in the framework of the Holstein electron-phonon model 
without additional including the anharmonic terms in the lattice potential. 

Recent investigations of Peierls transitions in electron-phonon systems 
have prompted us to study similar effects in several hydrogen-bonded 
solids. On the one hand, parallel sequences of plains (001) 
are formed in superionic state of M$_3$H(AO$_4$)$_2$ crystals. 
These hexagonal conducting plains consist of AO$_4$ groups connected by
virtual hydrogen bonds (see Fig.~1(a)). In the low-temperature phases
the frozen-in hydrogen bonds with only one index $f$ ($f=1,2,3$) form
well-defined sequences of dimers that involves the appearance of the 
parallel dimerized chain arrays consisting of ionic groups linked
by the $f$th hydrogen bond (see Fig.~1(b)).
\begin{figure}[htbp]
\epsfxsize=4.cm
\epsfysize=3.5cm
\centerline{\epsffile{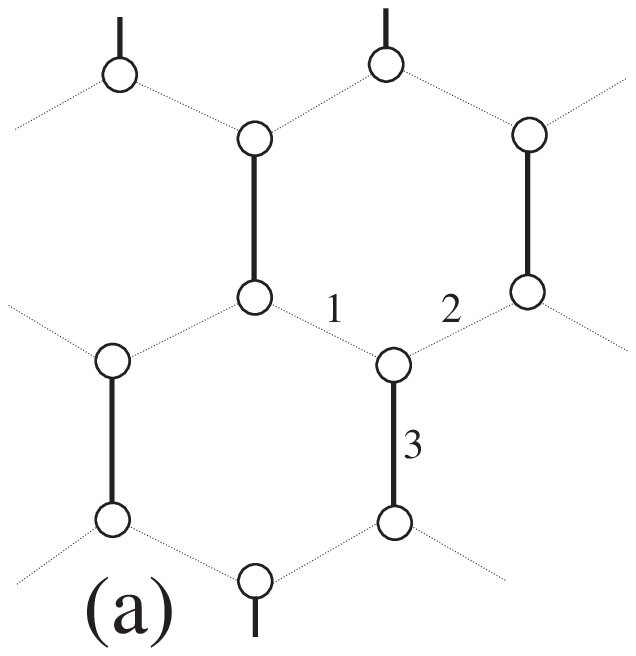}}
\epsfxsize=4.cm
\epsfysize=3.5cm
\centerline{\epsffile{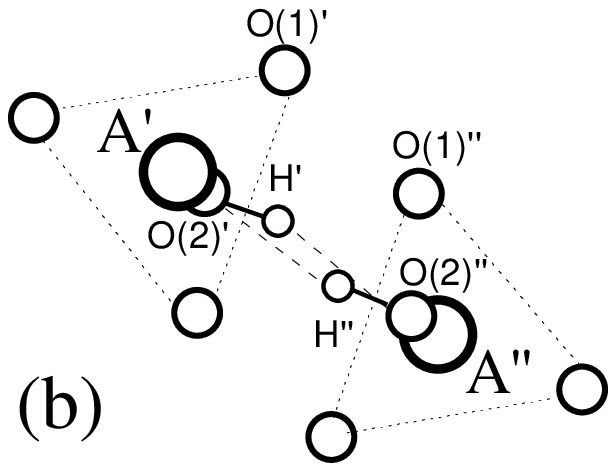}}
\caption{(a) Hydrogen-bonded network in (001) plane of M$_3$H(AO$_4$)$_2$
crystal group; the solid lines indicate the possible type of dimerized
structure which can appear with ($f=3$)th H-bonds frozen-in. 
(b) Structure of H-bonded dimer in one of low-temperature dimerized phases.}
\label{fig1}
\end{figure}
To analyze the influence of the 
proton-ionic group displacements coupling we consider the simplified model, 
namely the quasi-one-dimensional quantum double-well chain along one of the 
proton pathways (for instance, the virtual hydrogen bond sequence 
$\ldots-1-3-1-3-\ldots$). As an initial step, we neglect of the interproton 
repulsion effect that is justified for the low proton concentration (in our 
case each proton is averaged over every three virtual hydrogen 
bonds). However, we take into account the possibility of proton exchange 
between our selected chain and surrounding. On the other hand, besides 
these superionic compounds we also analyze in this work the influence of 
ionic group displacements on the proton subsystem behavior in 
quasi-one-dimensional solid hydrogen halides. We reveal possible 
symmetry-broken phases with proton charge disproportionalities coming from a 
Holstein coupling to AO$_4$ ionic groups or X atoms.  
We compare our conclusions with the results of the pressure effect 
theoretical studies in M$_3$H(AO$_4$)$_2$ \cite{sinitsyn} and hydrogen 
halides \cite{wang,jansen}. Although the first step of our 
analysis consists of the quasi-one-dimensional chain study, we believe our 
results can also be relevant for other hydrogen-bonded materials. 

\section{DESCRIPTION OF THE MODEL}

The object of our consideration is the chain shown in Fig.~2(a). 
However, to avoid the geometric complexities introduced by the kinks
in such an zig-zag chain, we consider in our model linear chain (see
Fig.~2(b) where two neighboring chains are shown). The process of the 
proton transfer in the double-well H-bond potential is represented as the 
quantum tunnelling between two proton states with intrabond transfer integral
$\Omega_0$
\begin{equation}
\Omega_0\sum_l(c_{la}^+ c_{lb}+c_{lb}^+ c_{la}), \label{h1}
\end{equation}
where $c_{l\nu}^+$, $c_{l\nu}$ denote proton creation and annihilation 
operators in the position ($l$, $\nu=a,b$) of the chain. Besides that, we 
describe the interbond reorientational proton hopping in two-level 
approximation as the quantum tunnelling effect with hopping amplitude 
$\Omega_R$
\begin{equation}
\Omega_R\sum_l(c_{l+1,a}^+ c_{lb}+c_{lb}^+ c_{l+1,a}).\label{h2}
\end{equation}
In this way in the framework of orientational-tunnelling model proposed in 
Ref.\ \onlinecite{jps} the two-stage proton migration 
mechanism can be considered 
as the sequential migration of the ionic and orientational defects.

As far as such a double-well chain is just a structural component of the 
system we also admit a possibility of proton exchange between the chain and 
surroundings by considering the system thermodynamics in the framework of 
the grand canonical ensemble with inclusion of the proton chemical potential
\begin{equation}
-\mu \sum_{l,\nu} n_{l\nu} \label{h3}
\end{equation}
which is to be determined at the given proton concentration in the chain 
from corresponding equation for the chemical potential. 
\begin{figure}[htbp]
\epsfxsize=9.cm
\epsfysize=3.cm
\centerline{\epsffile{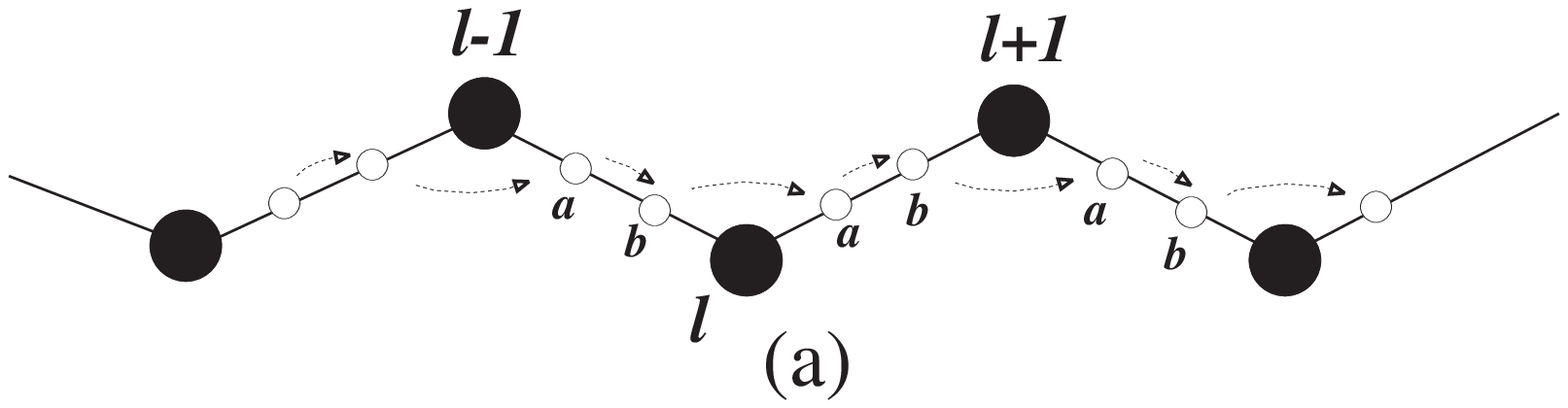}}
\epsfxsize=7.5cm
\epsfysize=3.cm
\centerline{\epsffile{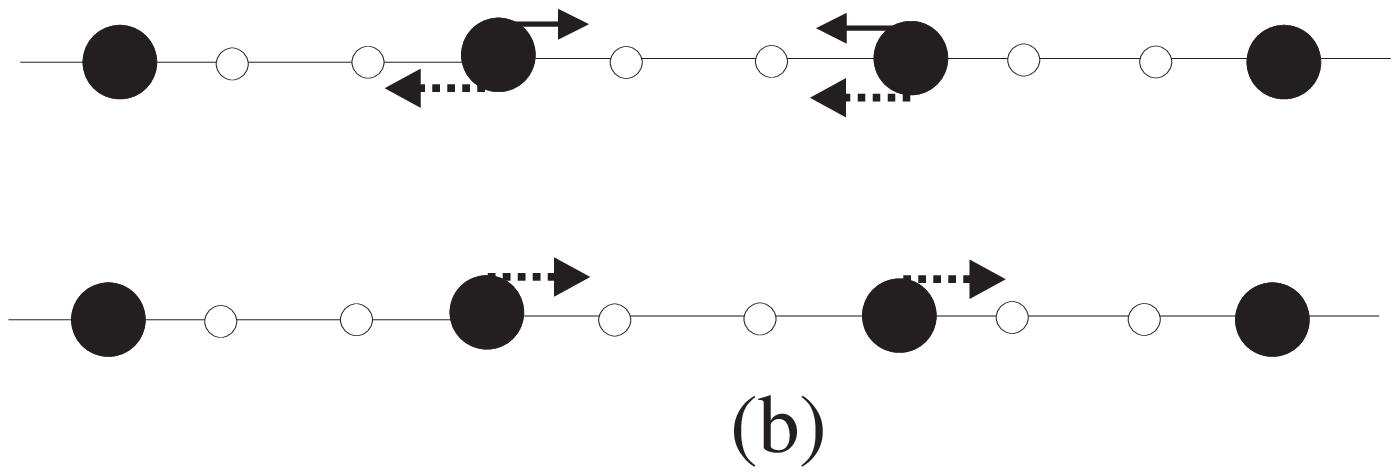}}
\null\vspace{0.1in}
\caption{(a) Zig-zag hydrogen-bonded chain in hydrogen halides, arrows
indicate the possible path of proton migration along the chain.
(b) Simplified model chains, the anti-phase and in-phase displacements
of ionic groups identified by solid and dashed arrows.}
\label{fig2}
\end{figure}

Our main interest is to analyze the influence of the longitudinal
optical ionic group vibration modes on the proton subsystem ground state. 
However, it was noted in Ref.~\onlinecite{springborg2} that the 
interactions between protons of the neighboring chains can lead to 
appearance of three-dimensional ordering. The more detailed analysis of the 
interchain proton interaction effect in this model together with
the determination of stability conditions for the existence of
the phases with different ordering type at finite temperatures will be 
presented elsewhere \cite{pavlenko2}.
We consider the anti-phase stretching vibration mode which causes a change 
of H-bond length in chain as indicated in Fig.~2(b) by solid arrows. 
Besides that, we also take into account the optical in-phase vibrations of
ionic groups in chain which induce their displacements with respect
to surrounding chains as identified in Fig.~2(b) by dashed arrows.
The coupling to the first type of displacements leads to the equal change of
the potential well ($l$, $a$) and ($l$, $b$) depth within the H-bond
\begin{equation}
\sum_{l,q} \tau_l^{(1)}(q)(n_{la}+n_{lb})(b_{q,1}+
b_{-q,1}^+), \label{h4}
\end{equation}
whereas the coupling of protons to another optical mode causes the 
difference of these potential minima depth
\begin{equation}
\sum_{l,q} \tau_l^{(2)}(q)(n_{la}-n_{l-1,b})(b_{q,2}+
b_{-q,2}^+). \label{h5}
\end{equation}
Here $\tau_l^{(1)}(q)=-2ig_1\sqrt{{\hbar}/{2MN\omega_1(q)}}
\sin\frac{1}{2}qd \exp[iq(l+1/2)d]$ and 
$\tau_l^{(2)}(q)=g_2\sqrt{{\hbar}/{2MN\omega_2(q)}}
\exp[iqld]$ where $g_1$ and $g_2$ are corresponding coupling 
constants, $M$ is the effective ionic group mass, $N$ denotes the number
of hydrogen bonds in chain and $d$ is a lattice 
spacing. Furthermore we take dispersionless approximation for the phonon 
frequencies: $\omega_1(q)=\omega_1$ and $\omega_2(q)=\omega_2$ and
assume the harmonic approximation for the 
lattice vibration energies
\begin{equation}
\hbar\omega_1 \sum_{q} b_{q,1}^+ b_{q,1}+ 
\hbar\omega_2 \sum_{q} b_{q,2}^+ b_{q,2}. \label{h6}
\end{equation}

First of all let us consider the case of the isolated chain without coupling 
to the phonon bath. Since the Hamiltonian (\ref{h1})-(\ref{h3}) can be 
exactly diagonalized, the proton energy spectrum 
\begin{equation}
\varepsilon_\nu (k)=\pm |t_{k}|, \hspace{0.05in}
|t_{k}|=\sqrt{\Omega_0^2+\Omega_R^2+2\Omega_0 \Omega_R \cos{kd}}
\end{equation}
forms two energy bands with the bandwidth 
$\Delta \varepsilon=\Omega_0+\Omega_R-|\Omega_0-\Omega_R|$. The energy gap
in this case is $\Delta_{ab}=2|\Omega_0-\Omega_R|$. Eliminating one of the 
elementary transport process by setting hopping amplitude $\Omega_0=0$
or $\Omega_R=0$ we can see that both the energy bands degenerate into two 
energy levels and the quantum fluctuations between these two system states 
could be derived. It is clear that in the case when 
$\bar{n}=\frac{1}{N} \sum\limits_{l\nu} \bar{n}_{l\nu}=1$ 
(one proton is averaged within 
the bond) the lower band is filled and the chemical potential $\mu$ is 
centered between bands - thus the material is in dielectric state. Such an 
situation can be observed in hydrogen halides. However, for 
$\bar{n}=\frac{1}{2}$ only half of the lower band is filled and this 
corresponds to the case of protonic conductor that occurs for example in 
superionic phases of superprotonic crystals.

We will discuss afterwards the consequences of the proton-phonon coupling 
effect focusing on the analysis of the two physically different cases 
$\bar{n}=\frac{1}{2}$ ($1/4$-filled two-band model) and $\bar{n}=1$
(half-filling case in two-band model).

\section{BROKEN-SYMMETRY SOLUTIONS}

\subsection{Case $\bar{n}=\frac{1}{2}$}

Let us now focus on the case of quarter filling when the half of the 
lower proton band is filled (one proton per two bonds). Then the 
macroscopic condensed phonon state is predominantly stabilized at 
$q^*=2k_F=\pi/d$ \cite{peierls,rise} and is characterized by the 
expectation values of the phonon creation and annihilation operators
\begin{equation}
\langle B_{q,1} \rangle=\langle b_{q,1}+b_{-q,1}^+ \rangle=
\frac{\Delta}{g_1} \sqrt{N} \delta_{q,q^*}, \label{d1}
\end{equation}
where $\Delta$ denotes the introduced distortion order parameter which 
should be determined from the stationarity conditions of the free energy. 
Since the condensation of displacements (\ref{d1}) leads to the unit cell 
doubling, using the Fourier transformation 
$c_{l\nu (i)}=\frac{1}{\sqrt{N/2}} \sum\limits_{k} c_{k\nu (i)} 
{\rm e}^{ikld}$ where the index
$i=\{+,- \}$ denotes ($l=2m$) or ($l=2m+1$)th cell,
the Hamiltonian in condensed state with static periodic distortions (\ref{d1})
(adiabatic treatment) is given by
\begin{eqnarray}
H=(-\mu+\tilde{\Delta}) \sum\limits_{k\nu} n_{k\nu (+)}- 
(\mu+\tilde{\Delta}) \sum\limits_{k\nu} n_{k\nu (-)}+ \nonumber\\
\frac{1}{8}N \frac{\tilde{\Delta}^2}{E_0}+
\Omega_0\sum_{k,i}(c_{ka(i)}^+ c_{kb(i)}+c_{kb(i)}^+ c_{ka(i)})+ \label{hc}\\
\Omega_R\sum\limits_k \sum\limits_{i\neq i'}
(c_{ka(i)}^+c_{kb(i')}{\rm e}^{-ikd}+c_{kb(i')}^+ c_{ka(i)}{\rm 
e}^{ikd}),\nonumber
\end{eqnarray}
where $E_0={(\hbar g_1)^2}/{2M (\hbar\omega_1)^2}$ is well known from
polaron theory \cite{polarons} protonic polaron binding energy which 
appears in the expression for the elastic energy per H-bond 
$\frac{1}{8}\frac{\tilde{\Delta}^2}{E_0}$ and
$\tilde{\Delta}=4\Delta \sqrt{\hbar/2M\omega_1}=4\Delta \sqrt{E_0 
\hbar \omega_1}/g_1$. The similar result can be obtained when we consider
the second type of the ionic group displacements, in this case
$\langle B_{q,2} \rangle=\langle b_{q,2}+b_{-q,2}^+ \rangle=
\frac{\Delta'}{g_2} \sqrt{N} \delta_{q,q^*}$ and the Hamiltonian
in condensed state has the form similar to (\ref{hc}) with 
$\tilde{\Delta} \rightarrow \tilde{\Delta}'=2\sqrt{\hbar/2M\omega_2}$
and $E_0 \rightarrow E_0'={(\hbar g_2)^2}/{2M (\hbar\omega_2)^2}$. 
Since the inclusion of the coupling to the second phonon mode leads merely
to renormalization of the binding energy $E_0$ in the Hamiltonian,
further we focus on the analysis of (\ref{hc}) with only one 
type of displacements taken into account. 
Introducing the double-time one-fermion diagonal Green functions
one can get rigorously the density of proton states
\begin{eqnarray}
&&\rho(\varepsilon)=\frac{2}{\pi} \frac{|\varepsilon| \cdot
|t_1-\varepsilon^2|}{B_1 B_2} \left\{ \Theta (\varepsilon \cdot {\rm 
sgn}(\varepsilon) -\sqrt{t_1-t_2^0})- \right.\nonumber\\
&&\Theta (\varepsilon \cdot {\rm sgn}(\varepsilon)-
\sqrt{(\Omega_0-\tilde{\Delta})^2+\Omega_R^2})+ \label{dos}\\
&& \Theta(\varepsilon \cdot {\rm 
sgn}(\varepsilon)-\sqrt{t_1+t_2^0})-\nonumber\\
&& \left. \Theta(\varepsilon \cdot {\rm sgn}(\varepsilon)-
\sqrt{(\Omega_0+\tilde{\Delta})^2+\Omega_R^2}) \right\}\nonumber
\end{eqnarray}
where
\begin{eqnarray}
B_1=\sqrt{(t_1-\varepsilon^2)^2-4\Omega_0^2\tilde{\Delta}^2},\\
B_2=\sqrt{4\Omega_0^2(\tilde{\Delta}^2+\Omega_R^2)-(t_1-\varepsilon^2)^2}
\nonumber
\end{eqnarray}
and the following notations are introduced: 
$t_1=\Omega_0^2+\Omega_R^2+\tilde{\Delta}^2$, 
$t_2^0=2\Omega_0\sqrt{\Omega_R^2+\tilde{\Delta}^2}$ and
$\Theta(x)$ is the Heaviside step function.
The expression for the ground state energy 
can be obtained easily from (\ref{hc}) and (\ref{dos}):
\begin{eqnarray}
F=\frac{N}{8}\frac{\tilde{\Delta}^2}{E_0}-\sum\limits_k \sqrt{t_1+2\Omega_0
\sqrt{\tilde{\Delta}^2+\Omega_R^2 \cos^2{kd}}}.   
\end{eqnarray}
To determine the stable phase the equation ${\partial F} / {\partial 
\tilde{\Delta}}=0$ should be solved.
It appears that this equations has besides $\tilde{\Delta}=0$,
a nonzero additional solution $\tilde{\Delta} \neq 0$ for 
$g_1>g_P$ where $g_P$ is the crossover proton-phonon coupling strength.
The solution $\tilde{\Delta}\neq 0$ corresponds to the global minimum
of $F$ and, as a result, implies the structural distortion stabilization 
with the amplitude $u_l=\sqrt{{\hbar}/{2MN\omega_1}} \langle 
B_{q^*} \rangle=\frac{\tilde{\Delta}}{2g_1} (-1)^l$ (see Fig.~3(a)).
Let us discuss the proton position average occupancies on the bonds
and the band structure. At $g_1=g_P$ each proton band splits into two 
subbands 
\begin{eqnarray}
\varepsilon_{a(+/-)}(k)=\mp 
\sqrt{t_1+2\Omega_0\sqrt{\tilde{\Delta}^2+\Omega_R^2 \cos^2{kd}}},\\ 
\varepsilon_{b(+/-)}(k)=\pm 
\sqrt{t_1-2\Omega_0\sqrt{\tilde{\Delta}^2+\Omega_R^2 \cos^2{kd}}}\nonumber 
\end{eqnarray}
as shown in Fig.~3(b) where the proton density of states
in the disordered and dimerized phases is represented.
The Peierls energy gap between either of the two
(lower and upper) subbands $\Delta_1=\sqrt{t_1+2\Omega_0\tilde{\Delta}}-
\sqrt{t_1-2\Omega_0\tilde{\Delta}} \approx {2\Omega_0\tilde{\Delta}}/
{\sqrt{\Omega_0^2+\Omega_R^2}}$ and tends to zero for $\Omega_0 \rightarrow 
0$. In this case $\tilde{\Delta}=\pm \sqrt{4E_0^2-\Omega_R^2}$ and the 
phase transition (change in the nature of the ground state)
occurs when the localization energy $E_0 \sim 
(g_1^*)^2=\frac{1}{2} \Omega_R$. The energy gap between the second and 
third subbands increases at $g_1>g_P$ 
\begin{eqnarray*}
\Delta_{ab}=2\sqrt{t_1-2\Omega_0\sqrt{\tilde{\Delta}^2+\Omega_R^2}}. 
\end{eqnarray*}
\begin{figure}[htbp]
\epsfxsize=5.5cm
\epsfysize=5.cm
\centerline{\epsffile{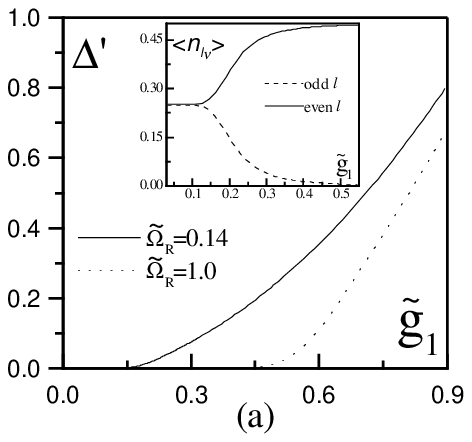}}
\null\vspace{0.1in}
\epsfxsize=5.5cm
\epsfysize=5.cm
\centerline{\epsffile{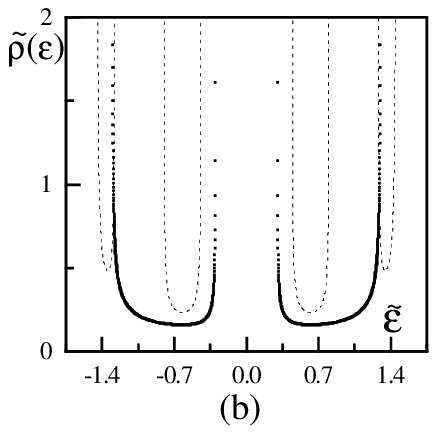}}
\null\vspace{0.1in}
\caption{(a) Distortion parameter 
$\Delta'=\frac{|\tilde{\Delta}|}{\hbar\omega_1}$
as a function of proton-phonon coupling 
$\tilde{g}_1$ for $\tilde{\Omega}_0=0.5$; inset: 
dependence of average proton accupancies on $\tilde{g}_1$ for 
$\tilde{\Omega}_R=0.14$. (b) Proton density of states 
$\tilde{\rho}(\varepsilon)=\frac{\rho(\varepsilon)}{\hbar\omega_1}$ 
($\tilde{\varepsilon}=\frac{\varepsilon}{\hbar\omega_1}$), 
dashed and dotted curves indicate the cases of $\Delta'=0.5$ and 
$\Delta'=0.0$ respectively.}
\label{fig3}
\end{figure}
The proton chemical potential $\mu$ is centered between two lowest 
subbands with further increasing of $g_1>g_P$ that points to the insulator 
state appearance. We see from inset in
Fig.~3(a) that the distortion stability is 
\begin{figure}[htbp]
\epsfxsize=8.cm
\epsfysize=1.cm
\centerline{\epsffile{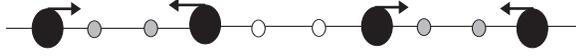}}
\caption{Dimerized structure which appears in the case of quarter-filled
chain.}
\label{fig4}
\end{figure}
accompanied by the formation of the proton charge-density-wave state in 
which $\langle n_{la}\rangle=\langle n_{lb} \rangle=\frac{1}{4}(1+(-1)^l)$ 
that means the forming of dimerized structure as shown in Fig.~4. Consider 
further the ground state phase diagrams 
($\tilde{g}_1={g_1}/{\hbar\omega_1}$, 
$\tilde{\Omega}_0={\Omega_0}/{\hbar\omega_1}$) and 
($\tilde{g}_1$, $\tilde{\Omega}_R={\Omega_R}/{\hbar\omega_1}$) 
represented in Fig.~5. We see the strong influence of the amplitude 
$\Omega_R$ on the dimerized state stability. The increasing of $\Omega_R$
suppresses dimerization. At $\Omega_R \rightarrow 0$ (without 
\begin{figure}[htbp]
\epsfxsize=4.5cm
\epsfysize=4.cm
\centerline{\epsffile{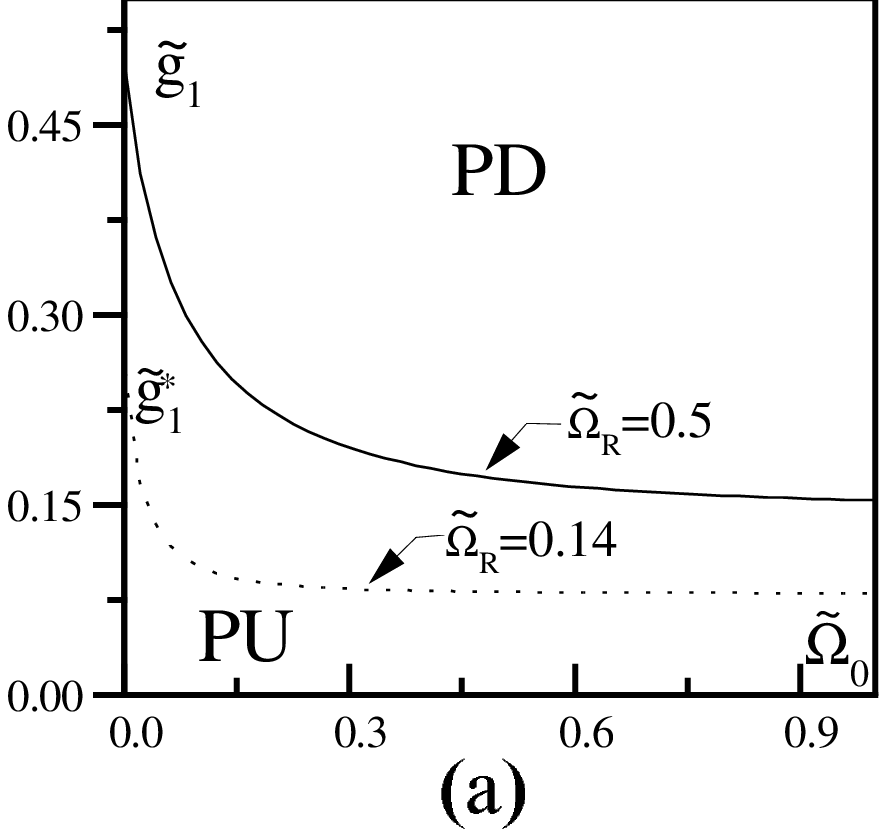}}
\null\vspace{0.1in}
\epsfxsize=4.5cm
\epsfysize=4.cm
\centerline{\epsffile{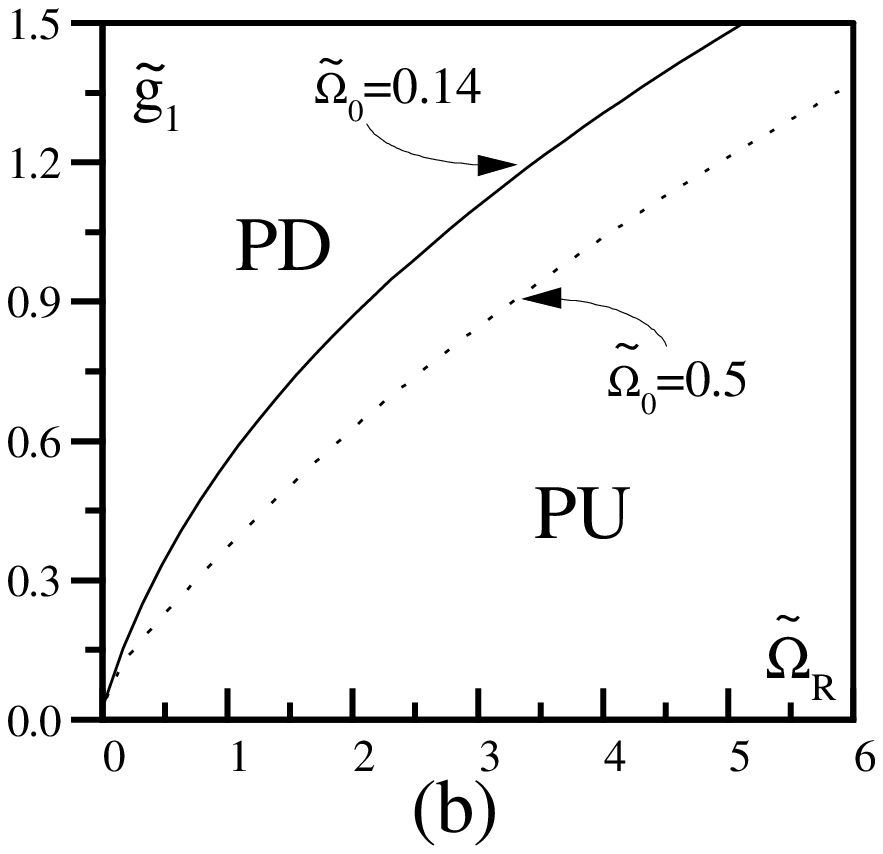}}
\null\vspace{0.1in}
\caption{Ground-state phase diagrams (a) ($\tilde{g}_1$, 
$\tilde{\Omega}_0$) and (b) ($\tilde{g}_1$, 
$\tilde{\Omega}_R$). The notations PD and PU denote the dimerized
and uniform phases respectively.}
\label{fig5}
\end{figure}
reorientational hopping) the system is brought immediately into the 
dimerized state. Only for the finite values of $\Omega_R$ the uniform 
disordered phase begins to appear and the "metal"-insulator transition 
occurs.

It is necessary to mention that the hopping amplitudes $\Omega_0$ 
and $\Omega_R$ depend strongly on external pressure. In particular,
the $\Omega_0$ value decreases with pressure that deduces from quantum 
mechanical calculations \cite{scheiner} as well as from the experimental 
measurements \cite{rambaud}. This is associated with the shortening of the 
distance between two potential minima ($l$, $\nu$) in the bond. Thus
we can make a conclusion about pressure effect on the system state from the
diagrams shown in Fig.~5. Using the obtained in \cite{pavlenko} values for 
parameters $\Omega_R$, $g_1$ and $\omega_1$ ($\Omega_R/\hbar\omega_1 
\approx 0.14$ and $\hbar^2 g_1^2/2M(\hbar\omega_1)^3 \approx 3.8$) we 
reveal that the dimerized state is always stable at T=0 under pressure for 
this set of parameters.
It is interesting that the similar picture has been 
observed in M$_3$H(AO$_4$)$_2$ materials from experimentally measured baric 
dependencies at low temperatures \cite{sinitsyn}. Nevertheless, we notice
that as $g_1$ decreases and approaches the critical value 
$g_1^*=\sqrt{\Omega_R/2}$, a transition from the dimerized to the uniform 
state occurs with pressure. This effect appears due to the more weak
proton-phonon coupling and, as a result, to the tendency of the proton
delocalization in chain.

\subsection{Case $\bar{n}=1$}

Let us discuss another case when one proton in average is placed in the 
bond. According to Peierls theory \cite{peierls} such a system is very 
susceptible towards lattice modulation at $q^*=0$. It should be noted that 
in this case only the second type of optical vibrations (interchain mode) 
contributes to the lattice distortions condensation. 
The Hamiltonian in condensed phase has the form
\begin{eqnarray}
H=&&\sum\limits_{k}\left[(-\mu+\tilde{\Delta})n_{ka}- 
(\mu+\tilde{\Delta}) n_{kb}\right]+ 
\frac{1}{8}N \frac{\tilde{\Delta}^2}{E_0}+\\
&&\sum_{k}(t_k c_{ka}^+ c_{kb}+t_k^* c_{kb}^+ c_{ka}).\nonumber
\end{eqnarray}
In this case the density of proton states 
\begin{eqnarray}
&&\rho(\varepsilon)=\frac{2}{\pi} \frac{|\varepsilon|}
{\sqrt{4\Omega_0^2\Omega_R^2-(\varepsilon^2-t_1)^2}} \times \\
&&\left(\Theta(\varepsilon \cdot {\rm 
sgn}(\varepsilon)-\sqrt{\tilde{\Delta}^2+
(\Omega_0-\Omega_R)^2})- \right.\nonumber\\
&& \left. \Theta(\varepsilon \cdot {\rm sgn}(\varepsilon)-
\sqrt{\tilde{\Delta}^2+(\Omega_0+\Omega_R)^2})\right) 
\nonumber
\end{eqnarray}
points to the two-band structure
\begin{equation}
\varepsilon_\nu (k)=\pm \sqrt{\tilde{\Delta}^2+|t_k|^2}
\end{equation}
with the Peierls energy gap $\Delta_{ab}=2\sqrt{\tilde{\Delta}^2+
(\Omega_0-\Omega_R)^2}$. The chemical potential always is centered between 
two bands, i.e. $\mu=0$.
We present the equation for determination of 
$\tilde{\Delta} \neq 0$ which follows from the stationary condition of 
$F$:
\begin{equation}
\frac{1}{4E_0}=\frac{1}{N}\sum_k 
\frac{1}{\sqrt{\tilde{\Delta}^2+|t_k|^2}}. \label{del2}
\end{equation}
The nonzero solution, which appears for $g_2>g_P$, corresponds to the
formation of a proton charge density wave in chain together with
the distortions $u_l=\tilde{\Delta}/2g_2$ stabilization (see Fig.~6).
\begin{figure}[htbp]
\epsfxsize=8.cm
\epsfysize=1.cm
\centerline{\epsffile{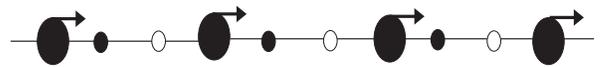}}
\caption{Broken-symmetry structure which appears in the case of 
half-filled chain.}
\label{fig6}
\end{figure}
The typical dependencies of the proton position average occupancies 
$\langle n_{l\nu} \rangle$ are represented in Fig.~7. It is interesting 
that the system now is invariable with respect to the interchanging 
$\Omega_0 \leftrightarrow \Omega_R$. Thus it is sufficiently to analyze the 
system behavior as a function of $\Omega_0$ for instance, with the given
fixed value of $\Omega_R$. The ground state phase diagram 
($\tilde{g}_2={g_2}/{\hbar\omega_2}$, 
$\tilde{\Omega}_0={\Omega_0}/{\hbar\omega_2}$)
(see Fig.~8) differs essentially from the case 
$\bar{n}=\frac{1}{2}$. The phase equilibrium curve has the specific
salient point at $\Omega_0=\Omega_R$. The drastically decrease of $g_P$
in the vicinity of $\Omega_0=\Omega_R$ is connected with the fact that the 
transfer anisotropy $|\Omega_0-\Omega_R|$ forms the additional transverse
field which competes with the ordering stabilization process. This 
anisotropy is vanished at $\Omega_0=\Omega_R$ that leads to the lowering of 
\begin{figure}[htbp]
\epsfxsize=6.cm
\epsfysize=6.cm
\centerline{\epsffile{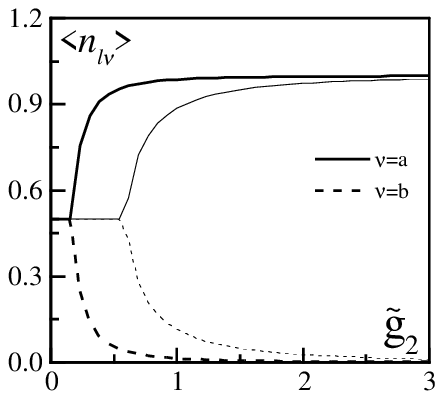}}
\caption{Average proton occupancies as a functions of $\tilde{g}_2$ for
$\tilde{\Omega}_0=0.8$; bold and thin curves indicate the cases when 
$\tilde{\Omega}_R=0.5$ and $\tilde{\Omega}_R=2.5$.}
\label{fig7}
\end{figure}
the crossover proton-phonon coupling energy $g_P$ required for the ordering 
stabilization. The interpretation of the diagram 
($\tilde{g}_2$, $\tilde{\Omega}_0$) with 
respect to the pressure effect is very interesting. 
\begin{figure}[htbp]
\epsfxsize=6.cm
\epsfysize=6.cm
\centerline{\epsffile{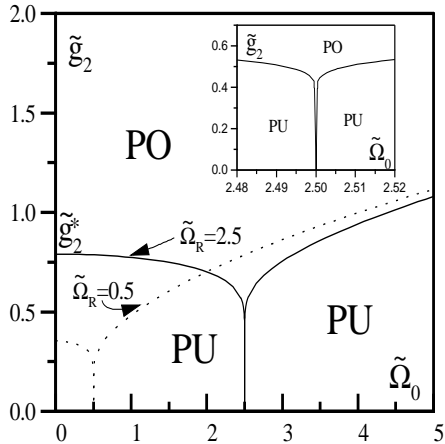}}
\caption{Ground-state phase diagrams ($\tilde{g}_2$, $\tilde{\Omega}_0$).
The notation PO indicates the symmetry-broken phase with proton ordering
on the hydrogen bonds; inset: the region $\Omega_0 \sim \Omega_R$ in more 
details.}
\label{fig8}
\end{figure}
The second-order transition from the uniform to ordered state occurs
under pressure (with $\Omega_0$ decrease) for $g_2>g_2^*=\sqrt{\Omega_R}/2$.
However, in the region $g_2<g_2^*$ the additional reentrant transition
from the symmetry-broken to the uniform state appears (see Fig.~9).
\begin{figure}[htbp]
\epsfxsize=6.cm
\epsfysize=6.cm
\centerline{\epsffile{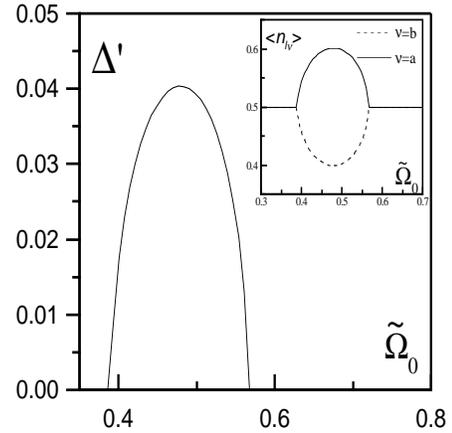}}
\caption{Distortion parameter as a function of $\tilde{\Omega}_R$
for $\tilde{g}_2=0.32$ and $\tilde{\Omega}_0=0.5$; inset: 
corresponding dependencies of average proton accupancies.}
\label{fig9}
\end{figure}
In this case the region of symmetry-broken phase equilibrium narrows
with $g_2$ decrease. We notice that the first-principle 
calculations \cite{jansen,springborg2,springborg}
and the results of Monte Carlo simulations \cite{wang} in quasi-one 
dimensional hydrogen halides show a transition from the symmetry-broken 
phase shown in Fig.~6 to the uniform symmetric phase under pressure at the 
low temperatures. Thus our results in the vicinity of 
$\Omega_0 \approx \Omega_R$
and for considerably weak proton-phonon coupling $g_2<g_2^*$
are in qualitative agreement with the conclusions of 
Refs.~\onlinecite{wang,jansen} 
confirming a proper treatment of the quantum effects in these 
hydrogen-bonded materials.

\section{CONCLUSIONS}
In the present work the lattice effect on the ground state properties
of the quantum quasi-one-dimensional hydrogen-bonded chain
is analyzed in the framework of the two-stage orientational-tunneling model.
The interaction of protons with two different types of surrounding ionic 
group optical displacements is considered. 
We show that when the proton-phonon coupling energy becomes large,
the system undergoes a transition from disordered to broken-symmetry
phases.
The different cases of proton concentration have been analyzed:
$\bar{n}=1/2$ and $\bar{n}=1$. It is shown that in the first case the 
Peierls transition to the dimerized phase occurs, whereas in the second 
one we obtain a transformation into the proton-ordered state. 
The influence of two different transport amplitudes on ground states 
properties is also studied. We compare 
our ground-state phase diagrams with the pressure effect experimental 
investigations in superprotonic systems and hydrogen halides at low 
temperatures.

\section*{Acknowledgements}

This work is partially supported by INTAS Grant No.~95-0133.

\onecolumn


\end{document}